\documentclass[prl,amsfonts,amssymb,floats,twocolumn,aps]{revtex4}

\usepackage[final,dvips]{epsfig}

\begin{document}

\title[Layer-Dependent Surface Magnetization]
{Layer-Dependent Magnetization at the Surface of a Band-Ferromagnet}

\author{R. Pfandzelter and M. Potthoff}

\affiliation{Humboldt-Universit\"at zu Berlin, Institut f\"ur Physik,
Invalidenstra\ss{}e 110, 10115 Berlin, Germany}

\begin{abstract}
The temperature-dependence of the magnetization near the surface
of a band-ferromagnet is measured with monolayer resolution. 
The simultaneous application of novel highly surface-sensitive 
techniques enables one to deduce the layer-dependent 
magnetization curves at a Fe(100) surface. 
Analysis of data is based on a simple mean-field approach. 
Implications for modern theories of itinerant-electron ferromagnetism 
are discussed.
\end{abstract}



\maketitle

A ferromagnetic material is characterized by a spontaneous
magnetization $m$ which decreases with increasing temperature $T$
until the paramagnetic state with $m=0$ is reached at the Curie
point $T_{c}$. For very low temperatures the form of the
magnetization curve $m(T)$ is governed by spin-wave excitations
according to Bloch's law. For temperatures very close to $T_{c}$
critical fluctuations result in a power-law dependence $m(T)
\propto (T_{c} - T)^\beta$ with a critical exponent $\beta$. In
the wide range of intermediate temperatures the form of $m(T)$
depends on the specific system. In case of a band-ferromagnetic
material, such as Fe as a prototype, the detailed form of $m(T)$
in this intermediate regime must be explained from the underlying
electronic structure \cite{aha96,BDN01}.

Density-functional theory within the local-spin-density approximation is
known to give a quantitatively accurate description of several ground-state
properties \cite{KV83}.
For finite temperatures, however, there is no satisfying implementation
available.
A microscopic theory must account for the existence of local magnetic
moments above $T_c$ in particular \cite{Mor85}.
This requires to deal with correlations among itinerant valence electrons
as, for example, within the framework of an orbitally degenerate Hubbard-type 
model with realistic parameters \cite{LKK01}. 
The long history of itinerant-electron ferromagnetism shows that this is 
a demanding task \cite{BDN01}. 
On the other hand, comparatively simple mean-field approaches based 
on spin models are known to provide a successful phenomenological 
description in many cases (see e.\ g.\ Ref.\ \cite{CL00}).
Remarkably, while the Weiss mean-field theory fails to reproduce the known 
$T \to 0$ and $T \to T_{c}$ limits and substantially overestimates $T_{c}$, 
the form of the Fe magnetization curve $m(T)$ at intermediate reduced 
temperatures $T/T_{c}$ is reasonably well described: 
For spin-quantum number $S=1/2$ there are deviations from the measured bulk 
magnetization curve of Fe within a few per cent only \cite{chi97,hah81}.

At the {\em surface} of a band-ferromagnet the magnetization may be different 
for different layers $\alpha$ parallel to the surface because of the reduced
translational symmetry.
Hence, a key quantity that characterizes the surface magnetic structure is
the layer-dependent magnetization curve $m_\alpha(T)$.
Within the framework of classical spin models, the lowered surface
coordination number implies that certain exchange interactions are
missing.
This directly leads to a reduced magnetic stability at the surface 
\cite{BH74}:
The top-layer ($\alpha=1$) magnetization is substantially reduced as
compared with the bulk.
However, significant deviations from the bulk magnetization curve are
confined to the first few layers in the intermediate temperature range.
This is confirmed qualitatively by calculations within Hubbard-type 
models of correlated itinerant electrons using slave-boson, decoupling 
and alloy-analogy approaches \cite{surf}.
Roughly, the results are similar to those for Ising or Heisenberg systems
in the range of intermediate temperatures.
Yet, the precise form of $m_\alpha(T)$ for a band-ferromagnetic surface 
must still be considered as largely unknown.


On the experimental side, determination of the layer-dependent
magnetization curve  at a surface is a demanding task as
well, which has not been achieved so far. In order to measure
$m_\alpha(T)$, experimental techniques sensitive to surface
magnetism are required with a magnetic probing depth tunable
with monolayer (ML) resolution. Common surface-sensitive
techniques like spin-resolved secondary-electron emission
\cite{ah87} or (inverse) photoemission \cite{ldd93} average over
several layers beneath the surface resulting in a (nearly)
bulk-like behavior of $m(T)$. Nevertheless, in a number of
sophisticated experiments a roughly linear temperature trend of
$m(T)$ has been observed and attributed to the surface
magnetization (see \cite{kgd84,ksg85,spr95} for Fe surfaces and 
\cite{DRW85} for experimental techniques).

Here we report on an experiment to determine
$m_\alpha(T)$ at the (100) surface of bcc Fe. The novel and
crucial feature of our experiment is the simultaneous application
of different {\it in-situ} techniques which are highly sensitive
to the magnetization near the surface, but slightly differ in
their magnetic probing depths. 

Ultimate surface sensitivity
(magnetic probing depth $\lambda = 0$~ML) is
achieved by spin-polarized electron capture \cite{whz89,ipw98}.
25~keV He$^{+}$-ions are grazingly scattered (incidence angle to
the surface plane $1 - 2^{\circ}$) off a magnetized Fe(100)
surface. The ions are reflected and capture target electrons into
excited atomic states. The spin-polarization of captured electrons
is deduced from the observed degree of circular polarization of
emitted fluorescence light. Excited atomic states can only survive
collisions for impact parameters exceeding the mean radius of the
corresponding electronic orbital. Thus the final
formation of atomic states takes place on the outgoing part of the
trajectories, resulting in a sensitivity of electron capture to a
region at or above the top surface layer.

An established technique to study magnetism near a surface is
spin-polarized secondary-electron emission, induced by keV
electrons at normal or oblique incidence \cite{s92}. Based on a
mean-field study, Abraham and Hopster \cite{ah87} infer from their
observed temperature-dependence of the spin-polarization of
secondary electrons from Ni(110) a magnetic probing depth of
$\lambda = 3 - 4$~ML (with an upper limit of 7~ML) for electrons
of about 0 or 10~eV kinetic energy. A compilation of electron
scattering cross sections by Sch\"onhense and Siegmann \cite{ss93}
suggests a similar value for Fe ($\lambda = 4.2$~ML), in
accordance with a recent overlayer experiment by Pfandzelter et
al.\ \cite{pow01}.

The probing depth in secondary-electron emission can be
considerably reduced by using energetic ions instead of electrons
as primary particles \cite{rwc90,pow01}. Grazingly incident ions
are reflected from the top surface layer and do not penetrate
into the bulk (``surface channeling''). In practice, structural
imperfections like surface steps mediate penetration of some
projectiles, leading to a contribution of excited electrons from
layers beneath the surface. From computer simulations emulating
ion trajectories \cite{p98} and an overlayer experiment
\cite{pow01}, we infer for scattering of 25~keV protons from our
Fe(100) surface a probing depth of $\lambda = (0.5 \pm 0.2)$~ML for
electrons of $10 - 20$~eV kinetic energy. We note that $\lambda$
seems to increase for lower electron energies owing to
cascade multiplication governed by electron-electron scattering
\cite{pow01}. Hence, energy resolution is mandatory if maximum
surface sensitivity is aspired.

Electron capture and electron emission yield information on the
spin part of the magnetization. 
Although a general quantitative relationship between
experimental observable (electron spin polarization) and
magnetization has not been worked out so far, it is generally
assumed that one can derive the (normalized) temperature
dependence of the magnetization. This assumption appears to be
justified in view of the, at least for the conditions of our
experiment, weak selectivity of capture and emission processes in
${\bf k}$-space.
Considering the small, well-defined, but different information
depths of the techniques, a simultaneous application at the same
surface thus should enable one to deduce the layer-dependent
magnetization curves near the surface.

\begin{figure}[t]
\centerline{\includegraphics[height=0.31\textwidth,angle=90]{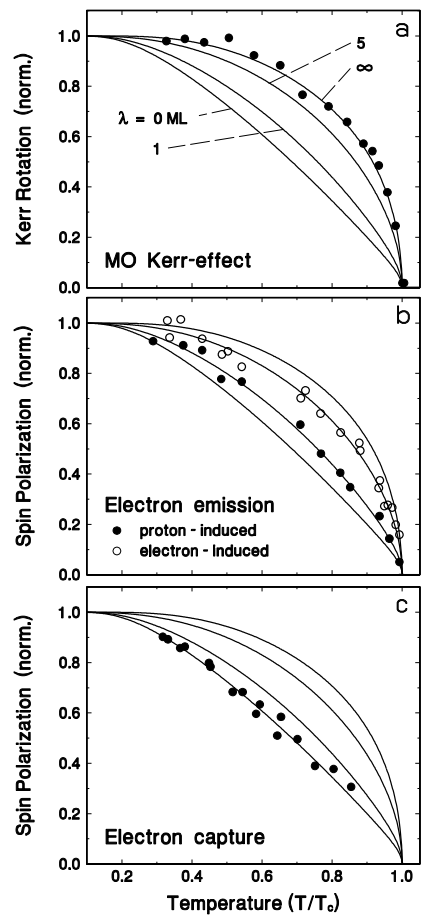}}
\caption{
Temperature dependence of (a)
Kerr-rotation, (b) spin-polarization of secondary electrons,
induced by electrons and grazingly scattered protons (open and
solid circles, respectively), and (c) spin-polarization of
electrons captured into atomic states of He$^{+}$-ions, grazingly
scattered off Fe(100). The curves represent mean-field
calculations according to Eqs.\ (\ref{eq:weiss}) and (\ref{eq:pol}) 
with different probing depths $\lambda$ as indicated.
}
\end{figure}

In our experiment, a (100) Fe single crystal disk
is mounted to close the gap of a magnetic yoke with a coil. For
the measurements the crystal is magnetized by current pulses
through the coil along an easy axis of magnetization $[001]$ or
$[00\bar1]$ in the (100) surface plane. This reproducibly yields a
full remanent magnetization near the center of the crystal as
checked by the magneto-optic Kerr effect. The (100) surface is
prepared by cycles of grazing Ar$^{+}$-sputtering and annealing,
until the surface is clean, atomically flat, and well-ordered, as
inferred from Auger electron spectroscopy, grazing ion-scattering,
and LEED. The target temperature is controlled by a thermocouple
attached directly near the crystal. Systematic differences between
the thermocouple reading and the crystal temperature are
calibrated by Kerr-effect (Curie temperature $T_{c}$) and
pyrometer measurements and corrected.

Electrons are emitted by 25~keV protons at grazing incidence
($1.2^{\circ}$) or by 4~keV electrons at oblique incidence
($33^{\circ}$) and enter an electrostatic energy analyzer
(cylindrical sector field) in a direction of about $10^{\circ}$
from normal. Spin analysis is performed for electrons with $10 -
20$~eV kinetic energy in a subsequent LEED spin polarization
detector \cite{pow01}. 
Each polarization spectrum is obtained from two
identical measurements with reversed magnetizations to eliminate
instrumental asymmetries.

Electron capture into the excited HeI 1s3p$^{3}$P-term formed
during grazing scattering of 25~keV He$^{+}$-ions is studied via
fluorescence light emitted in the 2s$^{3}$S-3p$^{3}$P, $\lambda =
388.9$~nm transition. The circular polarization fraction of the
light is measured by means of a rotatable quarter-wave plate, a
linear polarizer, a narrow band-width interference filter, and a
cooled photomultiplier. The transition being in the UV spectral
range, detection is affected on a tolerable level by stray light
from the filaments for heating the crystal up to temperatures
below about 900~K.

Experimental results are depicted in Fig.~1 as function of the
reduced temperature $T/T_{c}, \,  T_{c} = 1043$~K. The data are
collected from several quick heating and cooling runs. Short
measurement times turned out to be important in order to avoid
significant segregation of C at intermediate temperatures and S at
temperatures close to $T_{c}$ \cite{spr95}. No systematic
differences exceeding the statistical error (about $\pm 0.03$ for 
the normalized polarization) were
observed for heating and cooling runs, respectively. In Fig.~1~(a)
we show for comparison the rotation of the polarization axis
associated with the longitudinal magneto-optic Kerr-effect (solid
circles). The temperature-dependence is in agreement with previous
Kerr-rotation measurements at Fe(100) by Sirotti et al.
\cite{spr95} and reflects the bulk magnetization, because of the
large penetration depth of visible He-Ne-laser light (typically
20~nm \cite{abe88}).

Clearly different temperature-dependences are observed with the
surface-sensitive techniques electron-induced electron emission
(Fig.~1~(b), open circles), proton-induced electron emission
(Fig.~1~(b), solid circles), and electron capture (Fig.~1~(c),
solid circles). The curvatures gradually decrease, until, for
electron capture, an almost linear behavior is observed.
Remarkably, for electron-induced electron emission, a prominent
technique to study surface magnetism, the data closely resemble
the data from the Kerr-effect.

Information on the layer-dependent magnetization cannot be extracted 
from the data directly, as these have to be interpreted as
exponentially weighted averages over a number of layers
corresponding to the probing depth. We use an indirect way by
comparing with results of a simple mean-field calculation which is
known to reproduce the bulk magnetization curve fairly well,
provided that reduced quantities $m(T)/m(0)$ and $T/T_c$ are used.

Accordingly, the spin-$S$ Heisenberg model for the (100) surface of a bcc
lattice with layer-independent nearest-neighbor exchange $J$ is considered:
$H = - J \sum_{\langle i\alpha, j\beta \rangle}
{\bf S}_{i\alpha} {\bf S}_{j\beta}$.
Here $i$ labels the sites within a layer parallel to the surface and
$\alpha=1,...,\infty$ the different layers.
The mean-field free energy is
$F_{\rm MF} = - k_{\rm B}T \ln \mbox{tr} \exp({-H_{\rm MF}/k_{\rm B}T})$
where $H_{\rm MF}$ is obtained from $H$ by the usual decoupling
${\bf S}_{i\alpha} {\bf S}_{j\beta} \mapsto
\langle {\bf S}_{i\alpha} \rangle {\bf S}_{j\beta}
+ {\bf S}_{i\alpha} \langle {\bf S}_{j\beta} \rangle
- \langle {\bf S}_{i\alpha} {\bf S}_{j\beta} \rangle$.
Assuming collinear ferromagnetic order, ${\bf m}_\alpha = m_\alpha {\bf e}_z$,
and minimizing $F_{\rm MF}$ with respect to the order parameter,
${m}_\alpha = g \mu_{\rm B} \langle {S}_{i\alpha}^z \rangle$
($g$: Land\'e factor, $\mu_{\rm B}$: Bohr magneton),
yields a coupled set of Weiss self-consistency equations
\begin{equation}
  m_\alpha = m_\alpha(0) B_S ( S b_\alpha / k_{\rm B} T )
\label{eq:weiss}
\end{equation}
with $m_\alpha(0)=g\mu_{\rm B}S$, the Brillouin function $B_S$ 
\cite{Nol86b} and the layer-dependent Weiss field
${b}_\alpha = (2J/g\mu_{\rm B}) \left(
  z_\| {m}_\alpha + z_\perp {m}_{\alpha-1} + z_\perp {m}_{\alpha+1}
  \right)$. 
$z_\|=0$ and $z_\perp=4$ are the intra- and inter-layer coordination numbers
for the (100) surface. The total coordination number is $z=z_\| + 2 z_\perp = 8$.
The equations (\ref{eq:weiss}) are easily solved
numerically for a film of finite but sufficiently large thickness.
For the actual calculations we have taken $S=1$. Assuming the
orbital contribution to the magnetic moment to be quenched
completely ($g=2$), this appears to be the proper choice in the
case of Fe since the $T=0$ spin moment is $2.13 \, \mu_{\rm
B}$ per atom \cite{GR94}. 

To compare with the experiment we assume that each layer $\alpha$
gives a contribution proportional to $m_\alpha$ but weighted by an
exponential factor $\exp(-\alpha/\lambda)$ where $\lambda$ is the
probing depth characteristic for the experimental technique applied.
From the layer-dependent magnetization curves
$m_\alpha(T)$, we thus calculate a quantity $P(T)$ as
\begin{equation}
  P(T) \propto \sum_{\alpha=1}^\infty e^{-\alpha/\lambda}
  m_\alpha(T) \: .
\label{eq:pol}
\end{equation}
The results are shown in Fig.~1 (curves). The
mean-field $P(T)$ nicely reproduces the temperature trend of the 
(properly normalized) measured data for the respective information 
depth $\lambda$.
A value $\lambda = 5$~ML is consistent with the estimates 
for the probing depth in electron-induced electron emission
($\lambda = 4-5$~ML).
We have also checked against the choice $S=1/2$. This does not 
change the temperature trend of $m_\alpha / m_\alpha(0)$
as a function of $T/T_c$ significantly.
Surprisingly, considering an enhancement of the $T=0$ top-layer
magnetic moment (see Ref.\ \cite{OFW83}), does not lead to a
significant change of the temperature trend either.
Following Ref.\ \cite{SB88} one may expect a different exchange 
between the top- and the sub-surface-layer moments: $J_{12} \ne J$.
Within the experimental error, we find that the measured data are 
reproduced by calculations for a modified surface exchange in the
range from $J_{12}/J=0.8$ to $J_{12}/J=1.1$.

We conclude that the mean-field calculation gives a
rather accurate description of the layer-dependent magnetization at
the Fe(100) surface at intermediate temperatures. Clearly,
mean-field theory must be considered as a poor starting point to
explain surface magnetism. Nevertheless the result is interesting
as any theoretical approach that conceptually improves
upon the Weiss theory should give the same results (within our
experimental error).

In summary, this study for the first time gives detailed information 
on the layer-dependent magnetization at the surface of a prototypical
band-ferromagnet. We report on an experiment to measure 
temperature-dependent magnetization curves near the (100) surface of bcc
Fe. We simultaneously apply different techniques, two of which are
novel and based on grazing scattering of energetic ions, resulting
in an ultimate surface sensitivity. The magnetic information
depths of the techniques being well-defined but slightly different
enables one to achieve a near monolayer resolution.
The form of the layer-dependent magnetization
curve is an important key quantity of surface magnetism which, for
intermediate temperatures, represents a benchmark to discriminate
between different microscopic theoretical approaches to explain
surface magnetism from the underlying temperature-dependent
electronic structure.

\acknowledgments
The experimental part of this work was performed
in collaboration with T. Igel, M. Ostwald, and Prof. H. Winter.
Financial support by the Deutsche Forschungsgemeinschaft
(Sonderforschungsbereich 290) is gratefully acknowledged.


\begin{thebibliography}{99}

\bibitem{aha96}
A. Aharoni, {\em Introduction to the Theory of Ferromagnetism}
(Oxford University Press, Oxford, 1996).


\bibitem{BDN01}
K. Baberschke, M. Donath, and W. Nolting, eds., {\em
Band-ferromagnetism} (Springer, Berlin, in press).

\bibitem{KV83}
W. Kohn and P. Vashishta, {{\rm In:} \em Theory of the Inhomogeneous Electron Gas},
ed. by S. Lundqvist and N. H. March, p. 79 (Plenum, New York, 1983).


\bibitem{Mor85}
T. Moriya, {\em Spin Fluctuations in Itinerant Electron Magnetism},
  vol.~56 of {\em Springer Series in Solid-State Sciences}
 (Springer, Berlin, 1985).



\bibitem{LKK01}
A.I. Lichtenstein, M.I. Katsnelson, and G. Kotliar, preprint
cond-mat/0102297.

\bibitem{CL00}
R. E. Camley and Dongqi Li, Phys. Rev. Lett. \textbf{84}, 4709 (2000).

\bibitem{chi97}
S. Chikazumi, {\em Physics of Ferromagnetism} (Oxford University
Press, Oxford, 1997), p.118.

\bibitem{hah81}
S.D. Hanham, A.S. Arrott, and B. Heinrich, J. Appl. Phys.
\textbf{52}, 1941 (1981).

\bibitem{BH74}
K. Binder and P.C. Hohenberg, Phys. Rev. B \textbf{9}, 2194
(1974).

\bibitem{surf}
H. Hasegawa and F. Herman, J. Phys. (Paris) \textbf{49}, C8 1677 (1988);
M. Potthoff and W. Nolting, Surf. Sci. \textbf{377-379}, 457 (1997);
T. Herrmann and W. Nolting, J. Phys.: Condens. Matter \textbf{11}, 89 (1999).


\bibitem{ah87}
D.L. Abraham and H. Hopster, Phys. Rev. Lett. \textbf{58}, 1352
(1987).

\bibitem{ldd93}
W. von der Linden, M. Donath, and V. Dose, Phys. Rev. Lett.
\textbf{71}, 899 (1993).


\bibitem{kgd84}
J. Kirschner, M. Gl\"obl, V. Dose, and H. Scheidt, Phys. Rev.
Lett. \textbf{53}, 612 (1984).

\bibitem{ksg85}
E. Kisker, K. Schr\"oder, W. Gudat, and M. Campagna, Phys. Rev. B
\textbf{31}, 329 (1985).

\bibitem{spr95}
F. Sirotti, G. Panaccione, and G. Rossi, Phys. Rev. B \textbf{52},
R17~063 (1995).

\bibitem{DRW85}
F. B. Dunning, C. Rau, and G. K. Walters, Comments Solid State Phys.
\textbf{12}, 17 (1985).

\bibitem{whz89}
H. Winter, H. Hagedorn, R. Zimny, H. Nienhaus, and J. Kirschner,
Phys. Rev. Lett. \textbf{62}, 296 (1989).

\bibitem{ipw98}
T. Igel, R. Pfandzelter, and H. Winter, Phys. Rev. B \textbf{58},
2430 (1998).


\bibitem{s92}
H.C. Siegmann, J. Phys.: Condens. Matter \textbf{4}, 8395 (1992).

\bibitem{ss93}
G. Sch\"onhense and H.C. Siegmann, Ann. Physik \textbf{2}, 465
(1993).

\bibitem{pow01}
R. Pfandzelter, M. Ostwald, and H. Winter, Phys. Rev. B
\textbf{63}, 140406(R) (2001); R. Pfandzelter, M. Ostwald, and H.
Winter, Surf. Sci., in print.

\bibitem{rwc90}
C. Rau, K. Waters, and N. Chen, Phys. Rev. Lett. \textbf{64}, 1441
(1990).

\bibitem{p98}
R. Pfandzelter, Phys. Rev. B \textbf{57}, 15~496 (1998)


\bibitem{abe88}
J. Araya-Pochet, C.A. Ballentine, J.L. Erskine, Phys. Rev. B
\textbf{38}, 7846 (1988).



\bibitem{Nol86b}
W. Nolting, {\em Quantentheorie des Magnetismus}, vol.~2
(Teubner, Stuttgart, 1986).

\bibitem{GR94}
J.G. Gay and R. Richter, in: J.A.C. Bland and B. Heinrich (Eds.),
Ultrathin Magnetic Structures I (Springer, Berlin, 1994), p.21.

\bibitem{OFW83}
S. Ohnishi, A. J. Freeman and M. Weinert, Phys. Rev. B \textbf{28}, 6741 (1983).

\bibitem{SB88}
H. C. Siegmann and P. S. Bagus, Phys. Rev. B \textbf{38}, 10434 (1988).

\end{thebibliography}
\end{document}